\documentclass[12pt,aps,prb,superscriptaddress,preprint]{revtex4}   % style for Physical Review B and
\usepackage{amsmath}    % need for subequations
\usepackage{graphicx}   % for figures

\begin{document}

\title{Quantum machine using cold atoms}

\author{Alexey V. Ponomarev}
\affiliation{Institute of Physics, University of Augsburg,
Universit\"{a}tstr.~1, D-86159 Augsburg}
\email{alexey.ponomarev@physik.uni-augsburg.de}
\author{Sergey Denisov}
\affiliation{Institute of Physics, University of Augsburg,
Universit\"{a}tstr.~1, D-86159 Augsburg}
\author{Peter H\"{a}nggi}
\affiliation{Institute of Physics, University of Augsburg,
Universit\"{a}tstr.~1, D-86159 Augsburg} \affiliation{Department of
Physics and Center for Computational Science and Engineering,
National University of Singapore, 117542, Singapore}

\begin{abstract}
For a  machine to be useful in practice, it preferably has to meet
 two requirements: namely, (i) to be able to perform work under a load and (ii) its
operational regime should ideally not depend on the time at which
the machine is switched-on. We devise a minimal setup, consisting of
two atoms only,  for an ac-driven quantum motor which fulfills both
these conditions. Explicitly, the motor consists of two different
interacting atoms placed into a ring-shaped periodic optical
potential -- an optical ``bracelet" --, resulting from the interference
of two counter-propagating Laguerre-Gauss laser beams. This bracelet
is additionally threaded by a pulsating magnetic flux. While the
first atom plays a role of a quantum ``carrier", the second serves
as a quantum ``starter", which sets off the ``carrier" into a steady
rotational motion. For fixed zero-momentum initial conditions the
asymptotic carrier velocity saturates to a unique, nonzero value
which becomes increasingly  independent on the starting time with
increasing ``bracelet"-size. We identify the quantum mechanisms of
rectification and demonstrate that our quantum motor is able to
perform useful work.

\end{abstract}

\maketitle

\section{Introduction}

The development of laser cooling techniques and  the creation of the
Bose-Einstein condensates (BEC) present landmark successes of modern
experimental physics \cite{cooling}.  Since 1995 \cite{BEC},
ultracold atoms rapidly become a popular toolbox to explore the
quantum world. Subsequent experimental
studies can be (conditionally) divided into three stages. The first
one was aimed to model the electronic behavior in solids with
ultracold atoms trapped in periodic optical potentials. These
potentials, so called optical lattices thus implement the idea of
``quantum simulators", proposed by Feynman in 1982 \cite{Feynman}.
Incarnation of paradigmatic quantum models (Bose- and Fermi-Hubbard
systems, Tonk-Girardeau gas and etc) \cite{ober} and their intensive
experimental studies mark the second stage. The recent advent of a
third wave, namely ``to make use of it'', triggered a search for
applications which led to single atom qubits \cite{qubit}, atom
optical clocks \cite{clock}, cold atom interferometry \cite{inter},
as well as cold atom gyroscopes \cite{gyro}.

An electric motor, i.e.  a device that converts electrical energy
into mechanical work by setting a working body into a linear or
rotational motion, is a archetype example of a useful physical
system. For nearly two centuries, since the invention of the first
electrical motor \cite{first}, a never-ending continuous
miniaturization has already passed the microscale level \cite{MEMS}
and entered the nano-scale world \cite{nano1}. Yet, this process
alone does not parallel  the transition from the classical to the
quantum world: the  operational descriptions of all existing
electrical, nano-sized motors rest on classical concepts \cite{nano1}.

With this present work, we provide a detailed analysis for an
electric \textit{quantum} motor proposed recently in Ref \cite{qmotor}.
This  motor is made of two  ultracold atoms only that are trapped
in a deep ring-shaped one-dimensional optical lattice,
-- an optical ``bracelet". The blueprint for such an underlying
trapping potential was proposed recently \cite{amico} and a first
experimental realization has been reported in Ref \cite{ring_exp}.
We employ this setup to devise an engine which works as a genuine
ac-driven quantum motor. We identify the quantum mechanisms which
yield the {\it modus operandi} for a motor device. Moreover, we
discuss parameter values suitable for the realization of the
quantum engine with present-day experimental setups.

\section{The motor setup}

Figure 1 is a sketch of our quantum machine device. The optical
potential, which results either from the interference of a
Laguerre-Gauss (LG) laser beam with a plane wave \cite{amico} or,
alternatively, of two collinear LG beams with different frequencies
\cite{ring_exp} is capable of trapping two interacting atoms. The
first atom, termed ``carrier'', $c$, is assumed to be charged and is
driven by the time-dependent magnetic flux piercing the bracelet \cite{hub_rings}.
The second atom, termed ``starter'', $s$, is neutral thus remains
unaffected by the magnetic flux, but interacts locally, e.g. by
means of atom-s-wave scattering,  with the carrier when both atoms
share the same site of the optical lattice.

We next assume that both atoms are loaded into the lowest energy
band of a deep, ring-shaped optical potential with $L$ lattice sites
with the lattice constant $d$. The time-dependent homogeneous vector
potential $\tilde{A}(t)$ does not induce any appreciable transitions
between the ground band and the far separated excited band(s).

\begin{figure}[t]
\center
\includegraphics[width=0.6\textwidth]{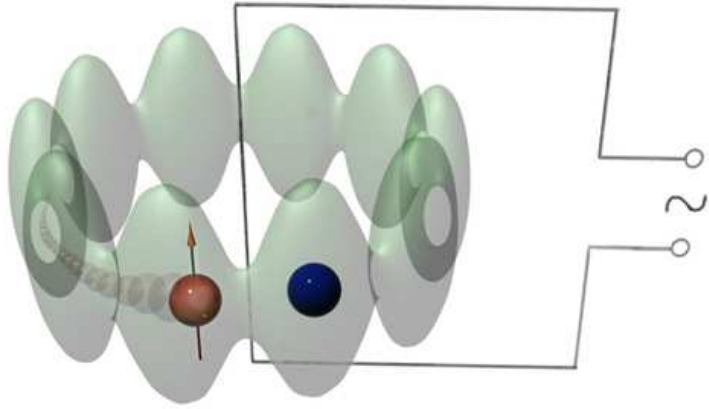}
\caption[Ring-shaped optical lattice] {\begin{small} Sketch
 of the quantum machine: Two ultracold neutral atoms are loaded into
an optical ``bracelet'' -- a ring-shaped optical lattice. Both atoms
interact locally with each other, while only one carrier (the one with
an arrow) is magnetically driven.
\end{small}} \label{fig1}
\end{figure}

The total Hamiltonian of the system reads
\begin{equation}
\label{eq:hamiltonian_total} H_{\rm tot} = H_{\rm c}(t) + H_{\rm s}
+ H_{\rm int}\,,
\end{equation}
where the time-dependent Hamiltonian $H_{\rm_c}$ for the carrier is
given by
\begin{equation}
\label{eq:hamiltonian_fermion}
H_{\rm c}(t) = -\frac{J_{\rm c}}{2}\left(\sum_{l_c=1}^Le^{i\tilde{A}(t)}|l_c+1\rangle_{\rm}\langle
l_c|_{\rm} + {\rm H.c.}\right)\otimes\mathbf{1}_s\,,
\end{equation}
and  the starter Hamiltonian $H_{\rm s}$ reads
\begin{equation}
\label{eq:hamiltonian_boson} H_{\rm s}= -\frac{J_{\rm
s}}{2}\left(\sum_{l_s=1}^L|l_s+1\rangle_{\rm}\langle l_s|_{\rm} +
{\rm H.c.}\right)\otimes\mathbf{1}_c\,.
\end{equation}
Here, $J_c$ and $J_{s}$ are the corresponding hopping strengths to
neigbouring sites, which depend on the atom mass, $M$, and the
potential depth, $V_0$. Here, for both atoms, we have assumed the
limit of a deep periodic potential, also refereed to as the
tight-binding model. This limit is well justified for a potential
amplitudes of $V_{0} \gtrsim 5E_0$, where the higher-order tunneling
amplitude is less than $10\%$ of the leading first-order (i.e., the
next neighbour) tunneling amplitude \cite{ober}.
$E_0=\hbar^2\pi^2/2Md^2$ is the ``recoil'' energy. The salient
carrier-starter on-site interaction is
\begin{equation}
\label{eq:interaction} H_{\rm int}=W \sum_{l_c,l_s=1}^L
\delta_{l_c,l_s}|l_c\rangle_{\rm}\langle l_c|_{\rm}\otimes
|l_s\rangle_{\rm}\langle l_s|_{\rm} \,,
\end{equation}
where $W$ denotes the interaction strength. Throughout the
remaining, we use periodic boundary conditions; i.e., $|L+1\rangle
=|1\rangle$. The dimension of the complete Hilbert space is
$\mathcal{N}=L^2$. The total system is described by wave function
$|\psi(t)\rangle$. The scale of the motor current will be measured
in units of the maximal group velocity $\upsilon_0= J_c d/\hbar$.

To conclude with the setup, we specify that the driving is switched
on at the time $t_0$, so that the vector potential has the form
\begin{equation}
\label{eq:switching_time} \tilde{A}(t;t_0) = {\mathit\chi(t-t_0)}A(t)\,.
\end{equation}
where ${\mathit\chi(t-t_0)}$ is the step function, and $A(t)$ is
defined on the entire time axis, $t \in (-\infty, +\infty)$.

\section{dc-quantum current}

The mean carrier current is given as the speed of the motor by means
of the velocity operator: $\hat
{\upsilon}_{c}(t;t_0)=i/\hbar\left[H_{\rm tot}(t),\hat x_c\right]$. With
$\hat x_c = d\sum_ll_c|l_c\rangle_{\rm}\langle l_c|_{\rm}\otimes\mathbf{1}_s$, one finds
$\hat {\upsilon}_{c}(t;t_0)=-i(\upsilon_0/2) \left(\sum_{l_c=1}^L
e^{i\tilde A(t;t_0)}|l_c+1\rangle_{\rm}\langle l_c|_{\rm}\right.$
$-\left.{\rm H.c.} \right)\otimes\mathbf{1}_s$. In the quasimomentum
representation with $|\kappa_l\rangle_{\rm} =
\sum_{n=1}^L\exp(i\kappa_ln)|n\rangle_{\rm}$, its quantum
expectation ${\upsilon}_{c}(t;t_0)=\langle\psi(t)|\hat
\upsilon_c(t;t_0)|\psi(t)\rangle$ reads
\begin{equation}
\label{eq:k_current} {\upsilon}_{c}(t;t_0) = \upsilon_0 \sum_{l=1}^L
\rho_{\kappa_l}(t;t_0) \sin\left(\kappa_l+\tilde{A}(t;t_0)\right)\,,
\end{equation}
wherein $\kappa_l = 2\pi l/L$ is the single particle quasimomentum
and where we indicated explicitly the parametric dependence on the
start time $t_0$. Further, $\rho_{\kappa}(t; t_0)  =
\sum_{l_s}|\langle\psi(t)|\kappa_{l_s},\kappa_{l_c}\rangle|^2$,
where
$|\kappa_{l_s},\kappa_{l_c}\rangle=|\kappa_{l_s}\rangle\otimes|\kappa_{l_c}\rangle$,
is the quasimomentum distribution for the carrier. The asymptotic
steady state  regime of the motor can be characterized by the
dc-component of the averaged velocity; i.e.,
\begin{eqnarray}
\label{eq:current_average_d} \upsilon_{c}(t_0) := {\rm
lim}_{t\rightarrow\infty}\frac{1}{t-t_0}\int_{t_0}^t\upsilon_{c}(s ; t_0)ds.
\end{eqnarray}

Without interaction between the particles, i.e., if $W = 0$, for an
initially localized carrier with zero velocity \textit{not even
a transient directed current does emerge} \cite{Hanggi1}. This
fact holds for \textit{any} shape of the potential $A(t)$. This
situation mimics the one taking place in a single-phase,
classical ac-motor: a periodically pulsating magnetic field would
fail to put a rotor from rest into rotation, unless one applies an
initial push via a starter mechanism \cite{motor}. Similarly, a
single quantum particle when initially localized on a single
potential minimum of a periodically modulated ring-shaped potential
doesn't acquire any momentum under an unbiased periodic driving
force \cite{Hanggi1}, while a constant bias, $A_{B}(t)=\omega_B
t$, induces Bloch oscillations only \cite{ober}. In our setup,
it is the ingredient of the interaction with the second particle
that takes over the role of a quantum starter.

Notably, even for nonvanishing interaction, i.e. $|W| > 0$, it is
still  not obvious how to set motor into rotation. Yet again, the
seemingly ``evident" solution -- to apply a constant bias to the
carrier -- cannot solve the task.
Due to the finite coupling  with the starter system in Eq.
(\ref{eq:hamiltonian_boson}), the corresponding vector potential,
$A_{B}(t)=\omega_B t$, may induce irregular Bloch oscillations with
a resulting zero drift velocity only \cite{Ponomarev}. In distinct
contrast, we use here an unbiased time-dependent vector potential
possessing a zero dc-component, $\int_{0}^{T}A(\tau)d\tau = 0$ and
being periodic in time, $A(t+T)=A(t)$.

Like for nanomechanical devices \cite{nano2}, the symmetry
principles are  of salient importance also  for quantum engines: for
a zero-momentum initial condition, an ac-force with time-reversal
symmetry would launch  the system -- with equal probabilities --
into a clockwise (rightward motion) or a counterclockwise rotation
(leftward motion) \cite{quantum}. Taking into account the quantum
nature of the engine, the rotor will just spread symmetrically in
both direction. Thus, the {\it modus operandi} requires a
symmetry-breaking driving field, realized here with the harmonic
mixing signal:
\begin{equation}
\label{eq:driving_vector_potential} A(t) = A_1\sin(\omega
 t)+A_2\sin(2\omega t+\Theta)\,.
\end{equation}
where $\Theta$ denotes the symmetry-breaking phase shift.
The input (\ref{eq:driving_vector_potential}) knowingly may induce a
non-vanishing nonlinear response, the so-called \textit{ratchet
effect} \cite{quantum, ratchet, Renzoni,goychukepl}.

\section{Floquet states as transport states}

The dynamics at times $t > t_0$ of the time-periodic Hamiltonian
(\ref{eq:hamiltonian_total}) can be analyzed by using the Floquet
formalism \cite{Hanggi2}. The solution of the eigenproblem:
$U(t,t_0)|\phi_n(t;k)\rangle =
\exp\left(-\frac{i}{\hbar}\epsilon_n(t-t_0)\right)|\phi_n(t;k)\rangle$,
with the propagator $U(t,t_0) = {\cal T}\exp
\left(-\frac{i}{\hbar}\int_{t_0}^t H_{\rm tot}(\tau)\right)d\tau$
(${\cal T}$ denotes the time ordering), provides the set of
time-periodic Floquet states, with $T= 2\pi/\omega$ being the
driving period, $|\phi_n (t+T;k)\rangle = |\phi_n(t;k)\rangle$.
Here, $k=2\pi l/L$ with $l=1,\dots, L$ is the {\it total} quasimomentum
of the Floquet state. Due to the discrete translation invariance of
the system, the total quasimomentum is conserved during the time
evolution, thus serving as a quantum number.

In the absence of the driving, $A(t)\equiv0$, the motor setup
(\ref{eq:hamiltonian_total}-\ref{eq:hamiltonian_boson}) possesses
the continuous translational symmetry in time. In this case, the
expansion coefficients of the initial wave-function $\psi(t_0)$ in
the system eigenbasis knowingly do not depend on time. On the contrary,
eigenstates of a periodically driven system -- the Floquet states --
evolve in time, being locked by the external ac-field. Thus, the
expansion of an initial wave-function over the Floquet eigenbasis
depends on the start time $t_0$ (\ref{eq:switching_time}), which
determines the phase of the driving ac-field \cite{quantum}, i.e.,
$|\psi(t_0)\rangle = \sum_{n=1}^\mathcal{N}c_n (t_0)|\phi_n (t_0; k)\rangle$,
with $c_n(t_0)=\langle\phi_n (t_0; k)|\psi(t_0)\rangle$.
Substitution of the above decomposition into (\ref{eq:current_average_d})
yields the result
\begin{eqnarray}
\label{eq:current_average} \upsilon_{\rm c}(t_0)
=\sum_{n=1}^{\mathcal{N}} \overline{\upsilon}_n\, |c_n(t_0)|^2 \;,\;
\overline{\upsilon}_n = \frac{1}{T} \int_{t_0}^{T+t_0} \upsilon_n(t)
dt\,.\;
\end{eqnarray}
Here, $\upsilon_n$ denotes the velocity expectation value
of the $n$-th Floquet state (\ref{eq:k_current}). Because the
Floquet states are periodic in time functions, the velocities
$\overline{\upsilon}_n$ do not depend on $t_0$, and the dependence
of the generated dc-current on the $t_0$ solely stems from the
coefficients~$c_n(t_0)$. Since the system evolution is fully quantum
coherent; i.e. there is no memory erasing induced by an environment,
-- the asymptotic current maintains the memory of the initial condition
as encoded in the coefficients $c_n(t_0)$ \cite{quantum}.

\section{Input/Output characteristics}
The question is now, how can we control the motor? To answer this
question, we used the symmetry analysis \cite{quantum} which allows
us to predict an appearance of a certain dc current. Combining
time-reversal operation and the complex conjugation applied to
(\ref{eq:hamiltonian_total}) with $A(t)$ in the form of, one can
prove the (anti-) symmetric dependence of $\overline{\upsilon}_n$ on
$\Theta$ for the Floquet states with $k=0$: (i)
$\overline{\upsilon}_n(-\Theta) = -\overline{\upsilon}_n(\Theta)$,
and (ii) $\overline{\upsilon}_n(\pi-\Theta) =
\overline{\upsilon}_n(\Theta)$. The first relation implies
$\overline{\upsilon}_n(0) = -\overline{\upsilon}_n(0) = 0$. Then,
for $\Theta=0$, the second relation gives
$\overline{\upsilon}_n(\pi) =0$. Thus the Floquet states with $k=0$
possess zero mean velocities at $\Theta=0,\pi$. Furthermore, using a
similar reasoning, one finds that the set of Floquet states with
nonzero $k$ can be ordered by the parity relation, which links
eigenstates with opposite quasimomenta, $\phi_n(t;-k;\Theta) =
\phi_m(T-t; k; -\Theta)$, yielding $\bar{\upsilon}_n(-\Theta) =
-\bar{\upsilon}_m(\Theta)$.  This implies that for a symmetric (in
$k$) initial state  and $\Theta=0,\pi$, the contributions to the
dc-current of Floquet states with opposite quasimomenta eliminate
each other. The same holds true for a monochromatic driving
(\ref{eq:driving_vector_potential}), with $A_2=0$ \cite{quantum}.
Shifting $\Theta$ away from  $0,\pm\pi$ causes the decisive symmetry
breaking and leads to the de-symmetrization of the Floquet states
with $k=0$ and consequently will violate the parity between states
with opposite signs of $k$.

\begin{figure}[t]
\center
\includegraphics[width=0.7\textwidth]{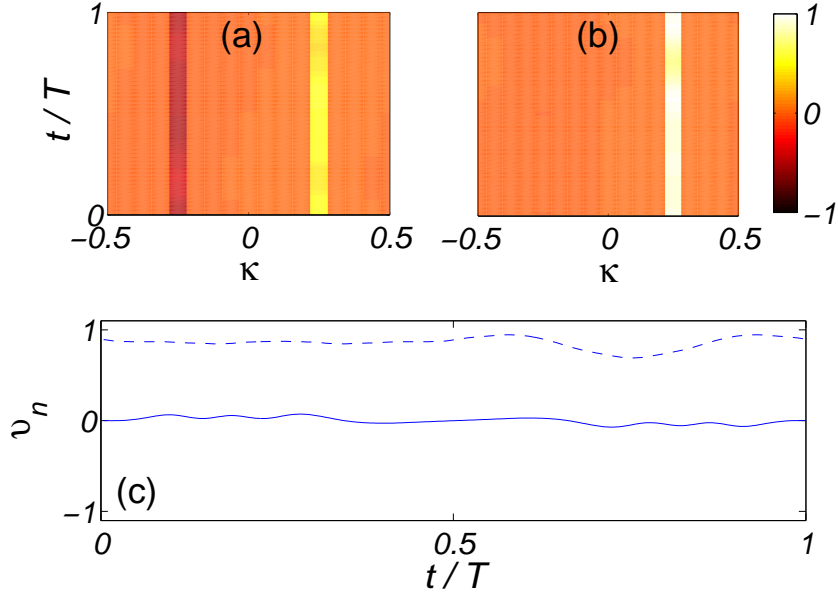}
\caption {(color online) The colormaps render velocities,
$\rho_\kappa(t;t_0) \sin\left(\kappa+\tilde{A}(t)\right)$, whose sum
according to (\ref{eq:k_current}) gives the mean velocity of an
individual Floquet state with the total quasimomenta $k=0$ for (a)
$\theta=0$ and (b) $\theta = \pi/2$, as a function of quasimomentum
$\kappa$. (c) Velocity of the Floquet states, $\upsilon_n(t)$, for
$\theta=0$ (solid line) and $\theta = \pi/2$ (dashed line). Note
that while $\upsilon(t)$ for $\Theta=0$ oscillates and is non-zero
except few points, its average over period, $\bar{\upsilon}_n$, is
strictly zero due to the symmetry $\upsilon_n(t)=-\upsilon_n(T-t)$.
The other parameters are $\hbar\omega = 0.1\times J_{\rm c}$, $A_1 =
0.5$, $A_2 = 0.25$, $W = 0.2 J_{\rm c}$, $J_{\rm s} = J_{\rm c} = J$,
$L=16$, and $t_0=0$.} \label{fig2}
\end{figure}

The emergence of a non-vanishing dc current at $\Theta\neq 0,\pi$
induced by the coupling to the starter can be illustrated by the
desymmetrization of velocities, $\rho_{\kappa}(t;t_0)sin\left(\kappa+\tilde A(t;t_0)\right)$
(as a function of quasimomentum $\kappa$), whose sum according to
(\ref{eq:k_current}) yields the mean velocity of the individual
Floquet states. For those with $k=0$, the desymmetrization happens
more drastically since they do not produce any dc current at the
symmetry point $\Theta=0, \pm \pi$. On Fig. \ref{fig2}, we depict
the instantaneous velocities, $\rho_{\kappa}(t;t_0)sin\left(\kappa+\tilde A(t;t_0)\right)$,
setting  $t_0=0$, for a Floquet state with $k=0$, for $\Theta=0$,
see Fig. \ref{fig2}(a), and $\Theta=\pi/2$, see Fig. \ref{fig2}(b),
together with the resulting mean velocities $\upsilon_n(t)$
Fig. \ref{fig2}(c). The contributions from different quasimomenta for
$\Theta=0$ eliminate each other and give a periodically varying,
but cycle-averaged zero  current, see in Fig. \ref{fig2}(c),
solid line, while the desymmetrization for $\Theta\neq 0$ (b)
results in non-vanishing dc current, cf. Fig. \ref{fig2}(c), dashed
line.

The motor speed depends on the initial conditions, which define the
contributions of different Floquet states to the carrier velocity
(\ref{eq:current_average}). We restrict our analysis to the initial
state $|\psi(t_0)\rangle
=L^{-1/2}|l_c\rangle\otimes\sum_{l_s}|l_s\rangle$, $l_c=1,...,L$, in
the form of the localized carrier (at $l_c$) and the uniformly
``smeared", delocalized starter. Both  particles assume zero
velocities at $t=t_0$. The asymptotic velocity may exhibit a strong
dependence on $t_0$ \cite{quantum}. We first discuss the results
obtained after averaging over $t_0$, thus assigning a unique motor
velocity value,
\begin{equation}
\label{eq:MC} \upsilon_c = \langle \upsilon_c(t_0) \rangle_{t_0} =
\frac{1}{T} \int_{t_{0}}^{T+t_{0}}\upsilon_c(t_0)dt_0 \;,
\end{equation}
for fixed system parameters.

\begin{figure}[t]
\center
\includegraphics[width=0.7\textwidth]{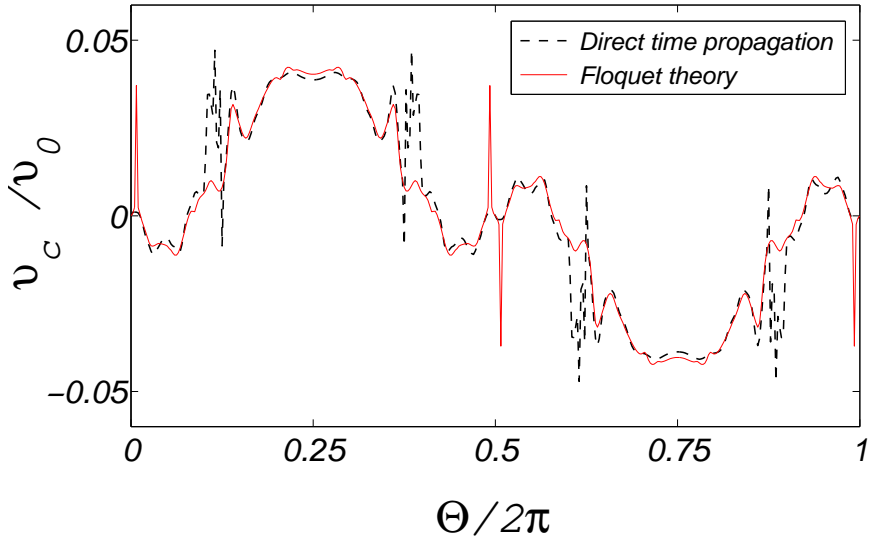}
\caption {(color online) Averaged motor velocity in (\ref{eq:MC})
(in units of the recoil velocity $\upsilon_0 = J_{\rm c}d/\hbar$) as
a function of the phase shift $\Theta$ in
(\ref{eq:driving_vector_potential}) for $L=16$. The ($t_0$)-averaged
velocity (\ref{eq:current_average_d}) obtained by the direct time
propagation of the initial state up to $200T$ (dashed line) is
compared to the asymptotic dependence given by the Floquet approach
(\ref{eq:current_average}) (red solid line). Note the anti-symmetry
behavior $\upsilon_c(\Theta)=-\upsilon_c(\Theta+\pi)$. The
parameters are the same as in Fig. 2.} \label{fig3}
\end{figure}

Figure \ref{fig3} depicts the dependence of the average motor
velocity on $\Theta$. The results obtained by direct time
propagation  of the initial state and averaged over $t_0$ (dashed
line) are superimposed by those calculated via the Floquet formalism
(\ref{eq:current_average}) (solid line). Both  curves show the
expected symmetry properties
$\upsilon_c(\Theta)=-\upsilon_c(\Theta+\pi)=-\upsilon_c(-\Theta)$.
The agreement between the two curves is satisfactory, although not
perfect: This is so because the sharp peaks on the asymptotic motor
velocity (\ref{eq:current_average}) and is due to the finite
evolution time used in direct time propagation.

\begin{figure}[t]
\center
\includegraphics[width=0.7\textwidth]{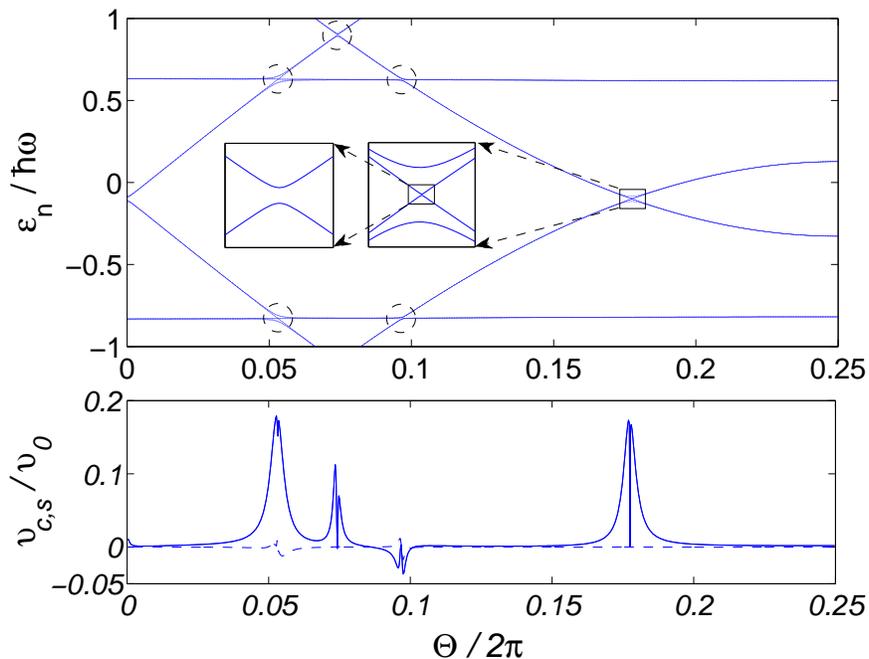}
\caption[Resonances of the asymptotic motor velocity] {\begin{small}
Resonances of the motor velocity. The quasienergy spectrum
(top) and asymptotic carrier/starter velocity (solid/dashed line)
in units of $\upsilon_0 = J_{\rm c}d/\hbar$ as a function
of $\theta$ (bottom), obtained numerically for $\hbar\omega = 0.05\times J_{\rm c}$,
$A_1 = 1$, $A_2 = 0.5$, $W = 0.01 \times J_{\rm c}$,
$J_{\rm s} = J_{\rm c}$, $L=4$. Presence of the avoided crossings
(emphasized by the circles and the rectangle frame) manifests
in the resonant velocity peaks at the corresponding values of
the phase $\theta$. The insets (top) zoom into the region of one of
the avoided crossing: the effect of a tiny anticrossing inside the
larger one is clearly resolved in the sharp resonance on top of the
wider one (bottom).
\end{small}} \label{fig4}
\end{figure}

The resonance peaks can be associated with \textit{avoided crossings}
between two quasienergy levels (see Fig. \ref{fig4}) \cite{quantum}.
These avoided crossings cause a strong velocity enhancement if one
of the interacting, and transporting eigenstate overlaps
significantly with an initial, non-transporting state of the motor.
Note also that a very narrow avoided crossing requires a very large
evolution time to become resolved, i.e., $t_{obs} \sim
\hbar/|\epsilon_{\alpha} - \epsilon_{\beta}|$, see in Ref
\cite{quantum}. Clearly, our chosen evolution time of $200T$ is
typically not large enough to clearly resolve the distinct
resonances depicted in Fig. \ref{fig2}.

With Fig. \ref{fig4}, we show an example of the quasienergy
spectrum, $\epsilon_n(\Theta)$, and the carrier and starter velocities,
$\upsilon_c$ (solid line) and $\upsilon_s$ (dashed line) respectively,
for the coupling constant $W/J=0.01$, and $L=4$. Here we remark that
a small coupling constant $W$ yields the avoided crossings smaller
while making the resonances sharper. This rather small system size,
however, provides already  a reasonable number of quasienergies for
our elucidation. For every velocity resonance depicted in Fig. \ref{fig4}
(bottom), one finds an avoided crossing in Fig. \ref{fig4} (top).
The fine avoided crossing structure induces an accompanying fine
structure of corresponding velocity resonances, note the two insets
in Fig. \ref{fig4}.

So far, we mainly focused on the motor velocity as given by the
carrier subsystem velocity. Let us here briefly also comment on
the starter dynamics, e.g., a possibly non-zero starter velocity.
We found that the averaged starter velocity $v_s$ indeed sensitively
depends on the system parameters: It can either be very small compared
to the carrier velocity (Fig. \ref{fig4}, bottom, dashed line) or also
larger than $v_c$. In short, the starter can move co-directionally
or contra-directionally to the carrier motion.

\begin{figure}[t]
\center
\includegraphics[width=0.7\textwidth]{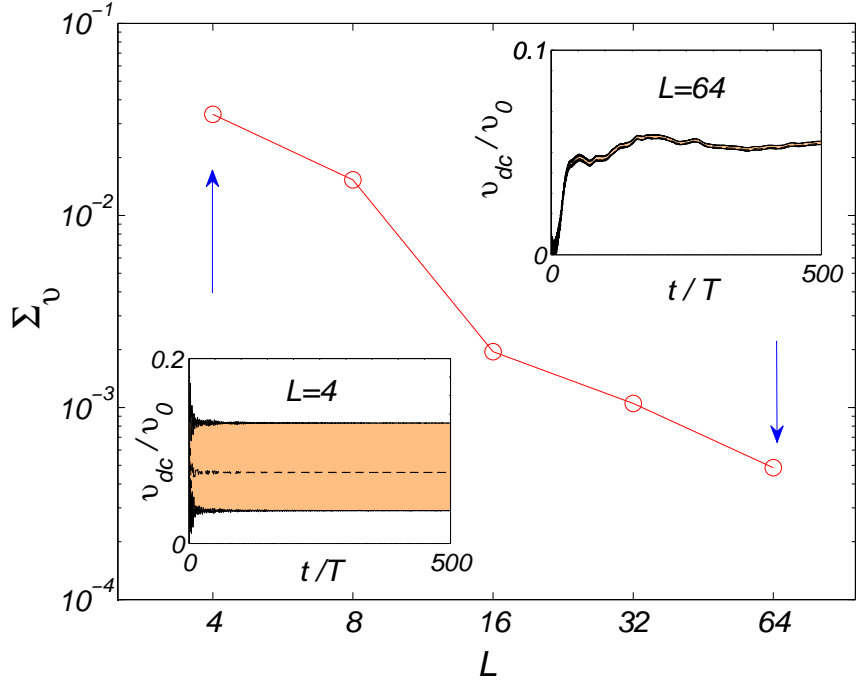}
\caption[Dispersion of the motor velocity.] {\begin{small}
Dispersion of the motor velocity (\ref{eq:dispersion}) \textit{vs.}
the number of lattice sites $L$. The dependence  of the asymptotic
direct current on $t_0$ vanishes with increase of the lattice size.
The insets show the running average for the carrier velocity. For
both insets  two realizations of $t_0$, which give the maximal and
minimal values of the asymptotic velocity are depicted (solid
lines), with remaining realizations for different $t_0$ varying in
between these lines. The dependence averaged over 20 realizations is
depicted in both insets by a dashed line. Note that for $L=64$ the
range bounded by two curves practically shrinks to the single thick
line. Here $\theta=\pi/2$, and the other parameters are the same as
in Fig. \ref{fig2}.
\end{small}} \label{fig5}
\end{figure}

A robustness of $\upsilon_{\rm c}(t_0)$ against a  variation of
$t_0$ characterizes the quality of our quantum motor, as it has been
mentioned in the abstract. Looking at the quasimomentum distribution
(similar to that of on Fig. \ref{fig2} (a,b)), it becomes evident
that $\rho_{\kappa}$ weekly depends on time, therefore we could
expect that the overlap with the initial state, $c_n(t_0)$,
according to (\ref{eq:current_average}), and as a result
$\upsilon_{\rm c}(t_0)$ has a week dependence on $t_0$. To provide a
more quantitative argument, we calculate the dispersion of the
asymptotic motor velocity with  respect to the switch on time $t_0$,
\begin{equation}
\label{eq:dispersion} \Sigma_\upsilon =
\sqrt{\left\langle\upsilon_{c}(t_0)^2-\langle\upsilon_{c}(t_0)\rangle_{t_0}^2
\right \rangle}_{t_0}\,.
\end{equation}
Here we use the direct numerical time-evolution over the
sufficiently long time, $t=500T$, to approach the asymptotic value
(\ref{eq:current_average_d}), see insets in Fig. \ref{fig5}). We
found that the dispersion becomes increasingly negligible with
increasing size $L$ of the lattice. Starting out from $L\gtrsim 16$,
the motor gains practically the same asymptotic velocity
independently on switch on time $t_0$.

This effect is due to the presence of the starter:  The carrier
velocity is obtained as the trace over the part of the total system
Hilbert space, $|l_c\rangle \otimes |l_s\rangle$, associated with
starter degrees of freedom, $|l_s\rangle$. The starter dynamics
mimics a finite ``heat bath'' for the carrier dynamics
whose effectiveness increases with both, the (i) the dimension of
the starter subspace, i.e. the size $L$, and (ii) the strength of
the interaction $W$.

\begin{figure}
\center
\includegraphics[width=0.7\textwidth]{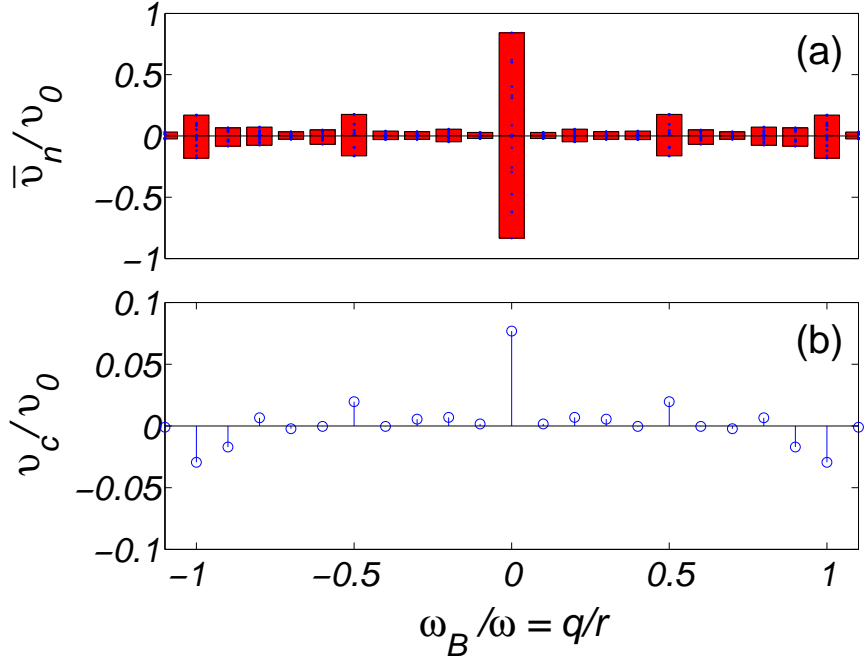}
\caption {(color online) Average motor velocity \textit{vs.} the
load $A_B(t) = \omega_B t$. (a) The range of velocity values for the 
Floquet eigenbasis. (b) The average motor velocity for the initial 
condition with the localized carrier and the delocalized starter. 
We set $\omega_B = \omega \cdot q/r$, with $r=10$, and vary the 
integer $q$. The parameters are $W=0.2 J_c$, $J_s = J_c= J$, 
$\hbar\omega = 0.1 J_c$, $A_1=0.5$, $A_2=0.25$, $\Theta=\pi/2$, 
and $L=4$.} \label{fig6}
\end{figure}

\section{Load characteristics} Thus far, the analysis of the system
in Eqs.(\ref{eq:hamiltonian_total} - \ref{eq:hamiltonian_boson},
\ref{eq:driving_vector_potential}) has been performed in a
idle-running mode with no load applied. In order to qualify for a
genuine motor device, the engine must be able to operate under an
applied load. The load is introduced in the form of an additional
bias $A_B(t)= \omega_Bt$, being added to the vector potential
$\tilde{A}(t)$. All the information about transport properties can
be extracted by using again the Floquet formalism, provided that the
ac-driving and the Bloch frequencies are mutually in resonance
\cite{kolovsky}; i.e. we have the condition  $q \cdot \omega = r
\cdot \omega_B$ obeyed, where $r$ and $q$ are co-prime integers.
Figure \ref{fig6} depicts the range of velocity values for Floquet 
eigenbasis, $\overline{\upsilon}_n$, see part (a), and the dependence 
of the resulting asymptotic motor speed, $\upsilon_c$, for different 
bias values, part (b). There occur two remarkable features. 
First, the spectrum of velocities is symmetric around $\omega_B=0$. 
This follows because of the specific choice of the phase shift 
at $\Theta=\pi/2$. Second, while some regimes provide a transport 
velocity along the bias, others correspond to the up-hill motion, 
against the bias. Therefore, a stationary transport in either 
direction is feasible. The load characteristics exhibits a discontinuous, 
fractal structure. In distinct contrast to the classical case 
\cite{kostur,kosturprb}, it cannot be approximated by a smooth curve. 
This is a direct consequence of the above mentioned resonance condition.

\section{Experimental realizations}

We next discuss the parameter range where the carrier atom generates
a tangible dc current. In the fast-driving limit, $\omega \gg
W/(\hbar(|A_1|+|A_2|))$, the Floquet states adiabatically follow
the instantaneous eigenstates of the total Hamiltonian in presence
of (\ref{eq:hamiltonian_total}) \textit{ zero particle-interaction}
i.e., $W=0$. Thus, the resulting, fast  driven dc current approaches
zero. In the slow-driving limit, $\omega \ll
W/(\hbar(|A_1|+|A_2|))$, the Floquet states adiabatically follow the
associated instantaneous levels of the static Hamiltonian
(\ref{eq:hamiltonian_total}) with no dynamical symmetry-breaking
field acting; thus the slow-driven dc current vanishes as well. Thus
the maximum dc current region is located in the intermediate region,
$\omega \sim W/(\hbar(|A_1|+|A_2|))$.

For an experimental realization of this quantum  atom motor the
following feature should be respected: Because in the tight-binding
approximation the maximal amplitude of the tunneling is limited from
above, $J_c \lesssim J_{max}=0.13 E_0$, for, e.g. a $^6\rm Li$
"carrier", a lattice constant $d\sim 10\mu m$, and with $\hbar
\omega = 0.1 J_c$ as used in the present calculations, the driving
frequency $\omega$ should be less than $2$Hz. Then, the time
required to launch the motor; i.e., for it to approach the
asymptotic velocity, which is $\sim 0.05 \upsilon_0 \approx 30\mu
m/s \approx 3$ sites/second, is around a minute. Further focusing of
the laser beam can decrease the lattice constant $d$, thereby
decreasing the launch time to experimentally accessible coherence
times around $10$ seconds \cite{gustavsson} and increasing the
asymptotic velocity.

\section{Conclusions}

We elaborated in greater detail on the working principles of the
quantum electric ac-motor that is made of two ultracold 
interacting atoms only \cite{qmotor}: a carrier and a starter, 
moving in a optical bracelet potential. The central result
of this study is an evident directed coherent carrier motion that
is induced by the starting mechanism. The  emerging motor velocity
can suitably be controlled by means of the symmetry breaking,
time-dependent, bichromatic, external magnetic flux.
Importantly, for zero-momentum initial conditions the asymptotic
carrier velocity loses its dependence on the switch-on time $t_0$
of the ac-drive upon increasing the bracelet size $L$.

An  extension of our motor setup to several interacting carries or
starters (i.e., multiple rotor motors or finite bosonic ``heat
baths'') presents an intriguing challenge. A particular interesting
objective to pursue is the problem of whether the motor velocity
can be optimally tuned with the number of participating atoms?

Finally, an exciting perspective is to physically couple our quantum
motor to a nano-mechanical resonator \cite{qm_resonators}. Such a
hybrid system can be used to power quantum mechanically such a
classical object.

This work was supported by the DFG through grant HA1517/31-1 and
by the German Excellence Initiative ``Nanosystems Initiative Munich (NIM)".

\end{document}